\RequirePackage{fix-cm}
\documentclass[twocolumn]{svjour3}          
\smartqed  
\usepackage{graphicx}
\begin{document}

\title{BCS-BEC crossover in quantum confined superconductors}

\author{Andrea Guidini         \and
        Luca Flammia		   \and
	    Milorad V. Milo\v{s}evi\'{c}\and        
        Andrea Perali
}

\institute{Andrea Guidini \at
              School of Science and Technology, Physics Division, University of Camerino, 62032 Camerino, Italy \\
              \email{andrea.guidini@unicam.it}           
           \and
           Luca Flammia \at
           School of Science and Technology, Physics Division, University of Camerino, 62032 Camerino, Italy\\
           Departement Fysica, Universiteit Antwerpen, Groenenborgerlaan 171, B-2020 Antwerpen, Belgium                      
           \and
			Milorad V. Milo\v{s}evi\'{c}\at
           Departement Fysica, Universiteit Antwerpen, Groenenborgerlaan 171, B-2020 Antwerpen, Belgium
           \and                      
           Andrea Perali \at
              School of Pharmacy, Physics Unit, University of Camerino, 62032 Camerino, Italy
}

\date{Received: date / Accepted: date}

\maketitle

\begin{abstract}
Ultranarrow superconductors are in the strong quantum confinement
regime with formation of multiple coherent condensates associated with the
many subbands of the electronic structure. Here we analyze the multiband
BCS-BEC crossover induced by the chemical potential tuned close to a subband
bottom, in correspondence of a superconducting shape resonance. The evolution
of the condensate fraction and of the pair correlation length in the ground
state as functions of the chemical potential demonstrates the tunability of
the BCS-BEC crossover for the condensate component of the selected subband.
The extension of the crossover regime increases when the pairing strength
and/or the characteristic energy of the interaction get larger. Our results
indicate the coexistence of large and small Cooper pairs in the crossover 
regime, leading to the optimal parameter configuration for high transition
temperature superconductivity.

\end{abstract}

\section{Introduction}
\label{introduction}
Control and enhancement of superconductivity when the material geometry
is reduced at the nano or atomic scale has been predicted long time ago.
First systems considered have been ultrathin superconducting slabs of metals
\cite{Blatt} and superconducting stripes \cite{Perali96}, with
the further extension to the superlattice of stripes, where the spatial
repetition of the stripes can stabilize macroscopic superconductivity 
against low dimensional fluctuations of the order parameter \cite{Bianconi97}.
Quantum size oscillations, shape resonances, and sizable amplification of the
superconducting critical temperature have been experimentally confirmed
and theoretically analyzed in metallic aluminum and tin nanowires
of cylindrical shape \cite{Shanenko2006}.
Interestingly, the Authors of Ref.\cite{Bianconi98} 
predicted a strong renormalization of the
chemical potential and the possibility of the complete condensation 
of all the electrons of the subband in which the chemical potential
is close to its energy bottom, giving rise to the coexistence of BCS-like
and BEC-like pairs. A more detailed study of the BCS-BEC
crossover in superconducting nanofilms induced by the quantum confinement
has been recently conducted in Ref.\cite{Chen2012}.

In another class of systems, ultracold cigar-shaped fermions, 
the quantum confinement induces the formation of a series of multiple subbands. 
In Ref.\cite{Shanenko2012} it has been found that in the  
superfluid state of these systems the overall condensate is a coherent 
mixture of partial subband condensates. Each partial condensate 
undergoes a BCS-BEC crossover when the chemical potential is tuned at the 
energy bottom of the corresponding subband. 

More recently, the discovery of the MgB$_2$ superconductor and of
the different families of iron-based superconductors 
definitely opened a new field of investigation in condensed matter physics:
the multicomponent (or multigap / multiband) 
superconductivity \cite{MMAP2015} (see also the web portal of the International
Network on Multi-component Superconductivity and Superfluidity: 
http://www.multisuper.org).
For iron-based superconductors the central role played by one of the bands
being in the BCS-BEC crossover was predicted in Ref.\cite{Caivano}.
Moreover, the presence of one band in the BCS-BEC crossover was predicted to
be a key feature for diverse high temperature superconductors in 
Ref.\cite{Bianconi2005}.
Recently, evidences of the BCS-BEC crossover 
in the small pockets of the Fermi surface of 
multi-band superconductors have been reported. ARPES in iron-calchogenide 
superconductors has reported a value of the ratio between the superconducting 
gap to the Fermi energy of order one in the shallow upper bands 
\cite{Kanigel,Terashima}. This phenomenology is now accepted to be a
generic important feature of all iron-based superconductors,
as discussed in Ref.\cite{BianconiNP2013}.
The observed phenomenology is the same as in ultracold 
fermions \cite{Perali2011,Pieri2011,Palestini2012}. 
A detailed discussion on these observations pointing to a component
selective BCS-BEC crossover in iron-based superconductors
and in related systems have been reported in \cite{Gui2014}. 
This is the optimal configuration for (very) high-T$_{\rm c}$ superconductivity. 
Multi-band BCS-BEC crossover makes possible
the screening of superconducting fluctuations, 
as shown in a two-band superconductor by a Ginzburg-Landau approach, 
see Ref.\cite{Perali2000}. 
In addition, multi-band BCS-BEC crossover is expected to induce an important
discrepancy of length scales of the different superconducting condensates
when the chemical potential approaches a band bottom of the multi-band system.
In this shape-resonant configuration, interesting vortex phenomena
with possible fractional states \cite{Geurts2010}, non-monotonic vortex
interaction \cite{Chaves2011}, and hidden criticality \cite{Komendova2012}
are predicted and could be experimentally observed.

Here we report on a simple physical realization of the 
multiband BCS-BEC crossover
induced by the quantum confinement in superconducting slabs at the atomic
limit. To capture the relevant and essential physics, we limit our analysis
to the first two subbands generated by the quantum confinement. We consider
an effective s-wave attractive interaction with an energy cutoff, including 
the important subband dependence induced by the reconfiguration of the
pairing at the atomic scale. By evaluating the superconducting 
ground state properties at a mean-field level of approximation, 
we obtain the selective BCS-BEC crossover when the chemical potential is 
tuned close to the second subband bottom.
The characterization of this crossover is obtained by evaluating the 
zero temperature condensate fraction of the partial condensate of the second 
subband and the correlation length of Cooper pairs forming in the 
same subband.

\section{Model and methods}
\label{model}
We consider a two dimensional two-band system of interacting fermions at $T=0$. The equations and relative discussions for band dispersion, gap, density, condensate fraction and pair correlation length are the same as discussed in Ref.\cite{Gui2014}. We here recall only the equations and the most important remarks.

The electronic subbands have a parabolic dispersion of the form
\begin{equation}
\label{dispersion}
\xi_i(\mathbf{k})=\frac{\mathbf{k}^2}{2m}-\mu +\varepsilon _i,
\end{equation}
where $m$ is the effective mass, which is taken equal in the two subbands, 
$\mu$ is the chemical potential and $\varepsilon_i$ are the 
energies of the subband bottoms. The index $i$=1, 2 labels the subbands: 
1 stands for the lower subband while 2 for the upper. 
Here $\varepsilon_1$ is set to zero.
The sketch of the subbands  and of the corresponding 
Fermi surfaces can be seen in Fig.~1 
of Ref.\cite{Gui2014}, with the difference that for the present work 
the picture presents the two-dimensional Fermi surfaces. 

The effective pairing attraction is taken in a separable form and with an 
energy cutoff $\omega_0$. In contrast with Ref.\cite{Gui2014}, here
the pairing interaction does not change structure when the chemical potential 
crosses the bottom of one of the subbands.
Moreover, because the motion along the $z$-axis 
is tightly bound, the bare strengths of the potential that control the 
intraband and the interband Josephson-like pairing between the two bands 
are related by $V^0_{ij}=V^0(1+\frac{1}{2}\delta_{ij})$. 
The pairing potential can be then written as
\begin{eqnarray}
\label{potential}
V_{ij}(\mathbf{k},\mathbf{k}')=-V^0(1+\frac{1}{2}\delta_{ij}) &\Theta&\left(\omega _0 - |\xi_i(\mathbf{k})|\right) \nonumber \\\times &\Theta&(\omega _0 - |\xi_j(\mathbf{k}')|),
\end{eqnarray}
where $V^0$ is the (positive) strength of the attractive (s-wave) potential.\\
The $\mathbf{k}$-dependence of the gaps is given by
\begin{equation}
\label{gaps_k}
\Delta _i(\mathbf{k})=\Delta _i \Theta(\omega _0 - |\xi _i(\mathbf{k})|).
\end{equation}

The coupled mean-field equations for the gaps are 
\begin{eqnarray}
\label{gap_eq_1}
\Delta _1(\mathbf{k})=&-&\frac{1}{\Omega} \sum_{\mathbf{k}'} \left[V_{11}(\mathbf{k},\mathbf{k}')\frac{\Delta _1 (\mathbf{k}')}{2\sqrt{\xi _1(\mathbf{k}')^2 +\Delta _1 (\mathbf{k}') ^2}}   \right. \nonumber \\ 
 &+& \left. V_{12}(\mathbf{k},\mathbf{k}')\frac{\Delta _2 (\mathbf{k}')}{2\sqrt{\xi _2(\mathbf{k}') ^2 +\Delta _2 (\mathbf{k}') ^2 }} \right],\\
\label{gap_eq_2}
\Delta _2(\mathbf{k})=&-& \frac{1}{\Omega} \sum_{\mathbf{k}'} \left[ V_{21}(\mathbf{k},\mathbf{k}')\frac{\Delta _1 (\mathbf{k}')}{2\sqrt{\xi _1(\mathbf{k}')^2+\Delta _1 (\mathbf{k}')^2}} \right. \nonumber \\ 
&+& \left. V_{22}(\mathbf{k},\mathbf{k}')\frac{\Delta _2 (\mathbf{k}')}{2\sqrt{\xi _2(\mathbf{k}') ^2 +\Delta _2 (\mathbf{k}')^2}}\right],
\end{eqnarray}
being $\Omega$ the area occupied by the system.

In this work the density is not fixed and it is given by
\begin{equation}
\label{density}
n_i=\frac{2}{\Omega}\sum_{\mathbf{k}}v_i(\mathbf{k})^2,
\end{equation}
where $v_i(\mathbf{k})$ is the BCS weight of the occupied states
\begin{equation}
\label{vk2}
v_i(\mathbf{k})^2=\frac{1}{2}\left[ 1-\frac{\xi_i(\mathbf{k})}{\sqrt{\xi_i(\mathbf{k})^2+\Delta _i(\mathbf{k})^2}}  \right].
\end{equation}
The condensate fraction $\alpha_i$ and pair correlation length $\xi_i$ are given by
\begin{eqnarray}
\label{condfrac_eq}
\alpha _i&=&\frac{\sum_{\mathbf{k}}(u_i(\mathbf{k})v_i(\mathbf{k}))^2}{\sum_{\mathbf{k}} v_i(\mathbf{k})^2}, \\
\label{csi_pair_eq}
\xi_i&=&\left[ \frac{\sum_{\mathbf{k}} |\nabla _{\mathbf{k}} (u_i(\mathbf{k})v_i(\mathbf{k}))|^2}{\sum_{\mathbf{k}} (u_i(\mathbf{k})v_i(\mathbf{k}))^2}\right]^{\frac{1}{2}}
\end{eqnarray}
where $u_i(\mathbf{k})^2=1-v_i(\mathbf{k})^2$.

The sums over $\mathbf{k}$ are replaced by two-dimensional integrals over momenta and then by integrals over the energy variable, after introducing the 2D density of states $N_{2D}=m/(2\pi)$.

The integrals of Eqs.~(\ref{gap_eq_1}-\ref{density}) and (\ref{condfrac_eq}) can be expressed in a closed form, while Eq.~(\ref{csi_pair_eq}) is calculated numerically. Indeed, to obtain the pair correlation length we introduce a smooth function in place of the step function associated to the $\mathbf{k}$-dependent gap of Eq.~(\ref{gaps_k}) in order to obtain well defined partial derivatives, and then we calculate numerically the integrals.

\section{Results}
\label{results}
We present here the plots of superconducting gaps, condensate fractions 
and pair correlation lengths as functions of the 
ratio $(\mu-\varepsilon_2)/\omega_0$, which is the Lifshitz parameter measuring
the distance from the Lifshitz transition in units of the energy cutoff of
the pairing interactions $\omega_0$,
for a given value of the cutoff energy $\omega_0/\varepsilon_2=0.5$ and for 
two couplings $\lambda \equiv N_{2D}V^0$=0.2 and 0.3. 
Gaps are units of $\omega_0$ as well as in units of the 
Fermi energies of the bands $E_{F_i}=\frac{2\pi}{m}n_i$ where $n_i$ is 
calculated by Eq.~(\ref{density}). The pair correlation lengths are 
measured in units of $1/k_{F_i}=1/\sqrt{4\pi n_i}$.

\begin{figure}
\includegraphics[scale=0.7]{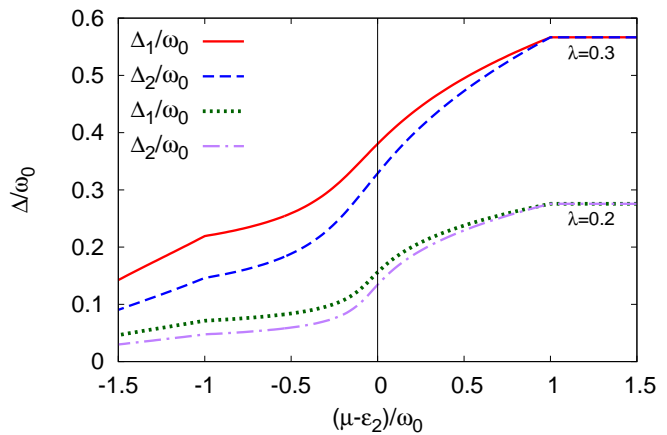}
\caption{Superconducting gaps $\Delta_{1,2}$ in units of $\omega_0$ as functions of the chemical potential $\mu$ in units of $\omega_0$ and referred to the bottom of the second subband $\varepsilon_2$ for $\omega_0/\varepsilon_2$=0.5 and two values of coupling $\lambda$=0.3, 0.2.}
\label{delta_su_eps}
\end{figure}
In Fig.~\ref{delta_su_eps} we present the superconducting gaps $\Delta_{1,2}$ in units of $\omega_0$ as functions of the chemical potential $\mu$ in units of $\omega_0$ and referred to the bottom of the second subband $\varepsilon_2$. The curves become flat when $\mu-\varepsilon_2-\omega_0 \geq 0$.

\begin{figure}
\includegraphics[scale=0.7]{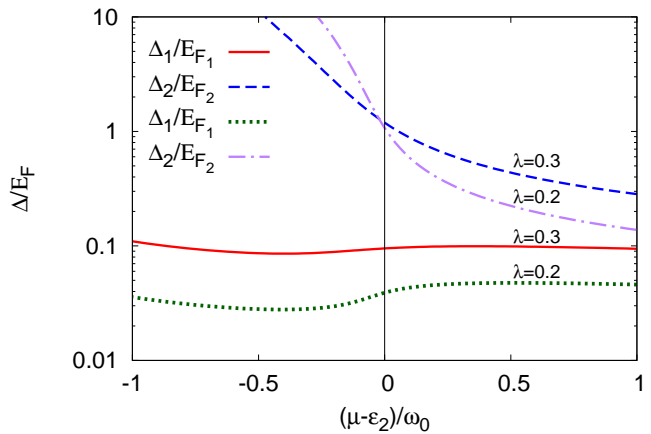}
\caption{Superconducting gaps $\Delta_{1,2}$ in units of the Fermi energies $E_{F_{1,2}}$ as functions of the chemical potential $\mu$ in units of $\omega_0$ and referred to the bottom of the second subband $\varepsilon_2$ for $\omega_0/\varepsilon_2$=0.5 and two values of coupling $\lambda$=0.3, 0.2.}
\label{delta_su_ef}
\end{figure}
In Fig.~\ref{delta_su_ef} we present the gaps in units of the Fermi energies of the subbands. The upper subband enters the BEC regime of pairing ($\Delta_2/E_{F_2}>1$) when the chemical potential crosses the bottom of the subband. In particular when $\mu \sim \varepsilon_2$ one has $\Delta_{2} \sim E_{F_{2}}$. The gap of the lower subband is instead less sensitive to $\mu$ and it signals that the pairing involving the lower subband is in the BCS regime.

\begin{figure}
\includegraphics[scale=0.7]{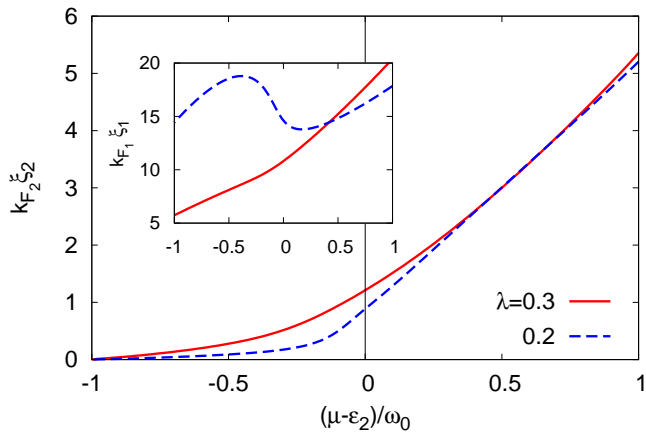}
\caption{Pair correlation lengths $\xi_{1,2}$ in units of $k_{F_{1,2}}^{-1}$ as functions of the chemical potential $\mu$ in units of $\omega_0$ and referred to the bottom of the second subband $\varepsilon_2$ for $\omega_0/\varepsilon_2$=0.5 and two values of coupling $\lambda$=0.3, 0.2.}
\label{csi}
\end{figure}
In Fig.~\ref{csi} we present the pair correlation lengths in units of the 
inverse of the Fermi wave-vectors of the bands. The evolution of the 
pair correlation length to describe the BCS-BEC crossover has been 
introduced in Ref.~\cite{Pistolesi94}. The lower subband is in the BCS 
regime ($k_F \xi \gg 1$), whereas the upper subband enters the 
BEC regime ($k_F \xi < 1$) when the chemical potential 
crosses $\varepsilon_2$. It is interesting also to note the feedback 
of $\xi_2$ on $\xi_1$ when $\mu=\varepsilon_2$ for $\lambda=0.2$. 
We found that this effect is more pronounced for small values 
of $\lambda$ and $\omega_0$.

\begin{figure}
\includegraphics[scale=0.7]{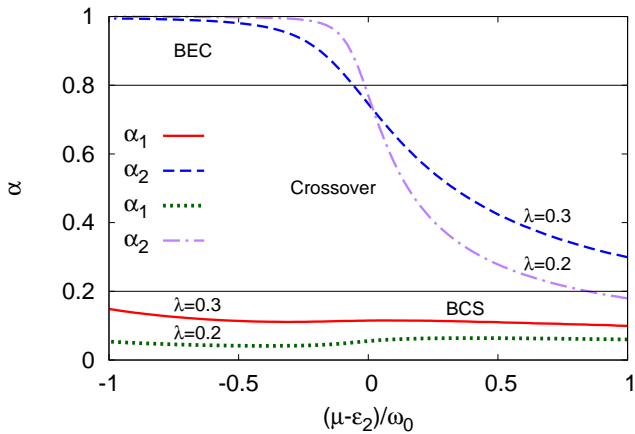}
\caption{Condensate fractions $\alpha_{1,2}$ as functions of the chemical potential $\mu$ in units of $\omega_0$ and referred to the bottom of the second subband $\varepsilon_2$ for $\omega_0/\varepsilon_2$=0.5 and two values of coupling $\lambda$=0.3, 0.2. The boundaries of the BCS-BEC crossover are indicated with horizontal lines.}
\label{alfa}
\end{figure}
In Fig.~\ref{alfa} we present the partial condensate fractions $\alpha_{1,2}$. 
Here we fix the boundaries of the different pairing regimes according to 
Ref.\cite{Gui2014}. The lower subband ($i=1$) is in the BCS regime of pairing 
having a condensate fraction smaller than 0.2 ($\alpha_{1}<0.2$) 
for every value of the chemical potential. 
The upper subband ($i=2$) is in the BEC regime, characterized
by a partial condensate fraction larger than 0.8 ($0.8<\alpha_{2}<1.0$), for 
$(\mu-\varepsilon_2)/\omega_0<0$ and in the crossover regime of the
BCS-BEC crossover
($0.2<\alpha_{2}<0.8$) for $0 < (\mu-\varepsilon_2)/\omega_0 < 1$. 
Larger values of the chemical potential ($(\mu-\varepsilon_2)/\omega_0 > 1$) 
will drive the system to a partial condensate fraction lower than 0.2 
also in the upper subband ($\alpha_{2}<0.2$). 
In this situation both the partial condensates 
of the two subbands are in the weakly-coupled BCS regime.

\section{Conclusions}
\label{conclusions}
Quantum confinement of superconductors at the atomic scale can 
generate a coherent overlap of partial superconducting condensates 
which are in different regimes of the BCS-BEC crossover.
We found that the ratio between the energy gap in one of the subbands and 
the local Fermi energy is the relevant experimental parameter to locate 
the partial condensate in the BCS-BEC crossover. 
Considering the second subband, 
when the chemical potential approaches the subband 
bottom the ratio $\Delta_2/E_{F_2}$ becomes of order 1 and the system enters 
the shape-resonant crossover regime when $0.3<\Delta_2/E_{F_2}<1.0$, 
with a rapid onset of the BEC regime for negative chemical potential 
with respect to the subband bottom, corresponding to $\Delta_2/E_{F_2}>1.0$.

\begin{acknowledgements}
We acknowledge A. Bianconi and A.A. Shanenko for useful discussions.
A.P. acknowledges financial support from 
the University of Camerino under the project FAR ``Control and enhancement of superconductivity by engineering materials at the nanoscale''. 
M.V.M. acknowledges support from the Research Foundation - Flanders (FWO) and the Special Research Funds of the University of Antwerp (BOF-UA).
A.P. and M.V.M. acknowledge the collaboration within the MultiSuper International Network (http://www.multisuper.org) for exchange
of ideas and suggestions.

\end{acknowledgements}

\end{document}